\documentclass[twocolumn]{aastex61}

\usepackage{amstext}
\usepackage{amsmath}
\usepackage{amssymb}
\usepackage{apjfonts}
\usepackage{graphicx}

\shorttitle{Accretion disks during CEE}

\shortauthors{Murguia-Berthier et al.}

\begin{document}

\title{Accretion Disk Assembly During Common Envelope Evolution: Implications for  Feedback and LIGO Binary Black Hole Formation}

\correspondingauthor{Ariadna Murguia-Berthier}
\email{armurgui@ucsc.edu}

\author{Ariadna Murguia-Berthier}
\affiliation{Department of Astronomy \& Astrophysics, University of California, Santa Cruz, CA 95064, USA}
\affiliation{Niels Bohr Institute, University of Copenhagen, Blegdamsvej 17, 2100 Copenhagen, Denmark}

\author{Morgan MacLeod}
\affiliation{School of Natural Sciences, Institute for Advanced Study, 1 Einstein Drive, Princeton, New Jersey 08540, USA}
\altaffiliation{NASA Einstein Fellow}

\author{Enrico Ramirez-Ruiz}
\affiliation{Department of Astronomy \& Astrophysics, University of California, Santa Cruz, CA 95064, USA}
\affiliation{Niels Bohr Institute, University of Copenhagen, Blegdamsvej 17, 2100 Copenhagen, Denmark}

\author{Andrea Antoni}
\affiliation{Department of Astronomy \& Astrophysics, University of California, Santa Cruz, CA 95064, USA}

\author{Phillip Macias}
\affiliation{Department of Astronomy \& Astrophysics, University of California, Santa Cruz, CA 95064, USA}
\affiliation{Niels Bohr Institute, University of Copenhagen, Blegdamsvej 17, 2100 Copenhagen, Denmark}

\begin{abstract}
During a common envelope episode in a binary system, the engulfed companion spirals to tighter orbital separations under the influence of drag from the surrounding envelope material.  As this object sweeps through material with a steep radial gradient of density, net angular momentum is introduced into the flow, potentially leading to the formation of an accretion disk.   The presence of a disk would have dramatic consequences for the outcome of the interaction because accretion might be accompanied by strong, polar outflows with enough energy to unbind the entire envelope. Without a detailed understanding of the necessary conditions for disk formation during common envelope, therefore, it is difficult  to accurately predict the population of merging compact binaries. This paper examines the conditions for disk formation  around objects embedded within common envelopes using the `wind tunnel' formalism developed by \citet{2017ApJ...838...56M}.  We find that the formation of disks  is highly dependent on the compressibility of the envelope material. Disks form only in the most compressible of stellar envelope gas, found in envelopes' outer layers in zones of partial ionization. These zones are largest in low-mass stellar envelopes, but comprise small portions of the envelope mass and radius in all cases. We conclude that disk formation and associated accretion feedback in common envelope is rare, and if it occurs, transitory. The implication for LIGO black hole binary assembly is that by avoiding strong accretion feedback, common envelope interactions should still result in the substantial orbital tightening needed to produce merging binaries. 
\end{abstract}

\section{Introduction}\label{int}

A common envelope (CE) phase develops in a binary system when one of the stars evolves off the main sequence and engulfs its companion \citep{1976IAUS...73...75P}.
Inside the CE an embedded object's orbit decays due to gravitational interaction with the surrounding gas. 
As orbital energy and momentum are exchanged with the CE gas, the envelope may gain sufficient energy and angular momentum to become unbound \citep{1984ApJ...277..355W, 1993PASP..105.1373I,2000A&A...360.1011N,2000A&A...360.1043D, 2000ARA&A..38..113T, 2010NewAR..54...65T, 2013A&ARv..21...59I}.

Depending on the efficacy of this envelope unbinding, the binary may either survive with a tightened orbit, or merge into a single object.  
The pathways  through which mass, angular momentum, and energy can flow through and around the CE thus play  a crucial role in establishing the outcomes of CE interactions and, more broadly, they determine the  imprint of CE on binary evolution \citep[e.g.][]{1993PASP..105.1373I,2013A&ARv..21...59I, 2014LRR....17....3P,2017PASA...34....1D}. These considerations are of particular importance when considering the assembly of compact objects into tight orbits  from which gravitational radiation can drive them to merger in less than a Hubble time \citep[e.g.][]{2002ApJ...572..407B,2007ApJ...662..504B,2007PhR...442...75K,2010ApJ...715L.138B, 2016Natur.534..512B}.

While the decay of the orbit is a known source of energy to the CE gas, there has also been discussion of whether accretion onto the embedded object could ``feedback'' and assist in unbinding the envelope gas \citep[see section 3.5 of ][for a discussion of this and other potential energy sources and sinks]{1993PASP..105.1373I}. 
Even the accretion of a small fraction of the CE mass onto a compact object could be sufficient to unbind the CE gas \citep[e.g.][]{2004NewA....9..399S,2015ApJ...800..114S}. For example, for an envelope of mass $M_{\rm env}$ with escape velocity $v_{\rm esc}$, an embedded black hole need only accrete a fraction $\Delta M / M_{\rm env} \gtrsim \left(v_{\rm esc} / c \right)^2$ to release sufficient energy to impinge upon or unbind the CE.
As a consequence, if accretion and associated feedback are major sources of energy in the CE event, the degree of orbital tightening required to eject a given CE (and terminate the interaction) would be drastically reduced.  A reduction in the orbital tightening experienced during the CE phase would, in turn, impact the population of compact binaries with merger times less than a Hubble time. If feedback from accretion were too efficient, we could imagine that CE-like interactions might produce no GW merger sources -- instead leaving behind only binaries too wide to merge today.

Answering these important questions  has not been straightforward, in large part because they depend on the details of the complex flow around objects embedded in CE. 
Gradients in the CE structure  introduce angular momentum into the flow about  the embedded object, potentially leading to the formation of a  rotationally-supported disk  \citep{2000ApJ...532..540A, 2013ApJ...767..135B}.  Disk structures, ubiquitous in astrophysical systems, create a mechanism through which these accretion-and-feedback flows persist: mass flows in the plane of the disk while energy is carried away vertically. In this case, inflow of mass, transported from large scales to an embedded companion, could be accompanied by prodigious mass loss driven by the outflows released by the accreted gas \citep[e.g.,][]{1991ApJ...376..214B, 1999MNRAS.303L...1B}. It's worth noting that the total outflow power need not be limited to the Eddington luminosity  \citep[e.g.][]{1980A&A....88...23P} as has, for example, been considered by \citet{2003MNRAS.342.1169V, 2016A&A...596A..58K}. 
 If these outflows were launched in the polar directions, they would impinge upon, and help unbind, material away from the binary orbital plane \citep{2000ApJ...532..540A, 2003MNRAS.342.1169V,2013arXiv1309.3925P, 2015ApJ...800..114S, 2017arXiv170203293M, 2017MNRAS.465L..54S}. 
Whether or not a disk forms may, therefore, have dramatic consequences on the accretion rate onto an embedded object and, also, on the accompanying feedback that could influence the CE gas at larger scales.  As will become clear, disk formation is particularly dependent on the thermal properties of the envelope, in particular, the response of the gas to compression.

To study the conditions under which disks can form in CE flows, we perform numerical simulations using the wind tunnel formalism developed by \citet{2015ApJ...803...41M} and \citet{2017ApJ...838...56M}. We explore local gas compressibility as a key parameter in shaping whether or not a disk forms around an embedded object. Section~\ref{num} introduces the numerical motivation and the formalism used. Section ~\ref{res} describes the results from our calculations. We will argue that disks form only in regions of high compressibility   with an adiabatic index $\gamma < 4/3$.
In Section~\ref{dis}, we extend our findings of the conditions under which disks are observed to form in order to study where these conditions are typically found in stellar envelopes.  We show that appreciable regions of sufficiently high compressibility occur in zones of partial ionization, and are likely only present in the envelopes of low mass giants. We argue that this implies that accretion feedback plays little role in shaping the outcomes of CE episodes involving binary black holes. And that, as a result, CE interactions with black holes should lead to substantial orbital tightening.

\section{Motivation and Numerical Formalism}\label{num}

\subsection{Background}

We will consider flow around a secondary object of mass $M_2$ and radius $R_2$ that is engulfed by its evolving companion (denoted here as the primary  star) with total mass $M_1$ and radius $R_1\gg R_2$. The pair has a mass ratio, $q=M_2/M_1$. The embedded object, separated by a distance $a\lesssim R_1$, will move within the CE with a characteristic  orbital velocity $v_{\rm k}^2(a)=G[M_{\rm 2}+M_1(a)]/a$, where $M_1(a)$ is the enclosed mass inside the orbit of the secondary. The orbital motion of the embedded object is likely to  be desynchronized from the  envelope of $M_1$ and the relative velocity can be written as $v_\infty = f_{\rm k}v_{\rm k}$, where $f_{\rm k}$ is the fraction of Keplerian velocity representing  the relative motion between the gas in $M_1$'s envelope and $M_2$. 

Studies of CE often  make use of Hoyle-Lyttleton accretion (HLA), a simple framework for understanding flow around an embedded secondary \citep[e.g.][]{1993PASP..105.1373I,2015ApJ...803...41M,2015ApJ...798L..19M, 2017ApJ...838...56M}. In this case, the object moves supersonically through the envelope and  gravitationally focuses the surrounding gas. Accretion is envisioned to take place if the impact parameter of the incoming gas is less than the accretion radius,
\begin{equation}
R_{\rm a}=\frac{2GM_2}{v_{\infty}^2},
\end{equation}
where $v_{\infty}$ is assumed to be supersonic \citep{1939PCPS...35..405H,1944MNRAS.104..273B,1952MNRAS.112..195B}. The corresponding mass accretion rate onto the embedded companion can then be written as
\begin{equation}
\dot{M}_{\rm HLA}=\pi R_{\rm a}^2 \rho_{\infty} v_{\infty},
\end{equation}
where $ \rho_{\infty}$ is the density of the incoming gas \citep[for a recent comprehensive review the reader is referred to][]{2004NewAR..48..843E}. The deflection of material  will result in a  reconfiguration of the flow, which in turn generates a net dynamical friction, drag  on the secondary \citep{1999ApJ...513..252O}.  

HLA was first  investigated numerically by \citet{1971MNRAS.154..141H} in order to determine whether   $\dot{M}_{\rm HLA}$  can provide an accurate estimate of the rate of mass accretion, concluding that it was indeed reasonable ($\dot{M}\approx0.88\dot{M}_{\rm HLA}$). This pioneering work laid the ground for several hydrodynamical studies for HLA  in  two \citep{1985MNRAS.217..367S, 2009ApJ...700...95B, 2013ApJ...767..135B} and three \citep{1994ApJ...427..342R,1994ApJ...427..351R,1994A&AS..106..505R,1995A&AS..113..133R,1996A&A...311..817R,2012ApJ...752...30B} dimensions. Of particular relevance to our work are the studies of \citet{1994ApJ...427..342R, 1994ApJ...427..351R, 1994A&AS..106..505R,1995A&AS..113..133R, 1996A&A...311..817R}, as they explored in great  detail the effects of varying  the properties of  the  background gas, in particular, the role of the  compressibility of the flow. If the flow is more compressible, the  loss of pressure support will result in a standing shock that resides closer to the accretor. The higher post-shock densities in addition to the steep pressure gradients, were shown to produce  higher mass accretion rates.  Simulations in two dimensions  showed that for highly  compressible gas, the  flow structure  becomes significantly  less stable, resulting in  large variations in the mass accretion rate \citep{2009ApJ...700...95B,2013ApJ...767..135B}. 

The HLA formalism has been widely used to describe the flow around objects embedded within a common-envelope,  but it fails to provide an accurate description of the flow. The formalism assumes a homogeneous background, which does not reflect the steep density profiles of evolving stars. Studies of HLA with vertical density and velocity gradients have been tackled by several groups \citep{1986MNRAS.218..593L,1986MNRAS.221..445S, 1986MNRAS.222..235L, 1987ApJ...315..536F, 1988ApJ...335..862F,1989ApJ...339..297T, 1997A&A...317..793R,1999A&A...346..861R,2000ApJ...532..540A,2015ApJ...803...41M,2015ApJ...798L..19M}, although in most cases the assumed  density gradients are shallow and are thus not representative of those found in stellar envelopes \citep{2015ApJ...803...41M}.  Symmetry breaking generated by the vertical gradient gives the flow net angular momentum relative to the accreting object. Even small gradients thus can have large-scale impacts on the flow, leading to rotational support for material, instead of radial infall as envisioned in HLA. 

The radial inflow approximation breaks down when the gas reaches a radius $R_{\rm circ}={l_z}^2/GM_2$, where $l_z$ is the specific  angular momentum  \citep[see, e.g., section 4.2 of][]{2015ApJ...803...41M}.  \citet{2000ApJ...532..540A} perform two-dimensional simulations with an  exponentially decreasing  density gradient  for  radiation pressure dominated  ($\gamma=4/3$) flows. They use a cylindrical geometry and assume a reflective inner boundary condition.  Under these conditions, they found  a stable centrifugally supported structure forming in their simulations. However, in recent  three-dimensional calculations using similar density gradients,   \citet{2015ApJ...803...41M} and  \citet{2017ApJ...838...56M} failed to produce rotationally-supported structures for radiation pressure dominated flows and only  saw disks when considering a softer equation of state.  

Generally, envelope gas may have a different response to compression under varying density and temperature conditions as determined by its equation of state.  
 The thermodynamic  description of the flow can be characterized by adiabatic exponents, 
\begin{equation}
\gamma_1= \left(\frac{d \ln P}{d \ln \rho} \right)_{\rm ad},
\end{equation} 
and,
\begin{equation}
\gamma_3=1+ \left(\frac{d \ln T}{d \ln \rho} \right)_{\rm ad},
\end{equation}
where the subscript signals  partial derivatives along a particular adiabat. $\gamma_1$ is relevant for calculating the  sound speed of the gas, $c_{\rm s}^2=\gamma_1 P/\rho$ while $\gamma_3$ is related to the  
equation of state, 
\begin{equation}\label{eos}
P=(\gamma_3-1)\rho e,
\end{equation}
where $e$ is the internal energy. In general  $\gamma_1$ will be greater than $\gamma_3$ when radiation plays a prominent role because in that case pressure increases  faster than temperature in response to compression. 

Material in a disk dissipates its motion perpendicular to the orbital plane, forming a differentially rotating structure. A net flow of material inward results when a viscosity-like stress transports angular momentum content outwards in the shear flow. A dynamo process of some kind is commonly believed to work and simple physical considerations suggest that fields generated in this way would have a length-scale of the order of the disk thickness and could  drive a  strong hydromagnetic wind \citep{1991ApJ...376..214B}. As discussed in the introduction, if the embedded object is a compact object,  this outflow could have enough kinetic energy to substantially alter the structure of the envelope.  

In the remainder of this work, we discuss how the properties of the stellar envelope have a decisive effect on whether or not a rotationally-supported structure can form  around an object embedded within a CE. 

\subsection{Model and Numerical Setup}

We perform idealized simulations of the flow around an embedded object using the CE Wind Tunnel (CEWT) formalism presented by \citet{2017ApJ...838...56M}. The inviscid hydrodynamic equations are solved using FLASH \citep{2000ApJS..131..273F}, an Eulerian, adaptive mesh refinement code.  
This setup models the  embedded secondary companion as a sink point particle of radius $R_{\rm s}$ at the origin. 
A wind, representing the gaseous envelope with a vertical profile of density and pressure, is fed in the $+x$-direction from the $-x$ boundary. This profile is in hydrostatic equilibrium with a vertical ($\hat y$), external $y^{-2}$ gravitational acceleration representing the gravity of the enclosed mass of the primary star. 
The code and methodology are described in detail in \citet{2017ApJ...838...56M}, but we include a few key points here for context. 

\subsubsection{Local Description of the CE}
The vertical profile of density and pressure within the envelope are locally approximated with a polytropic profile of a massless envelope (which assumes that the enclosed mass is small compared to the total mass of the primary across the region simulated $\sim R_{\rm a}$). In this case, the pressure and
density profiles of the surrounding envelope are described by 
\begin{equation}
\frac{d\rho}{dr}=-g\frac{\rho^2}{\Gamma_{\rm s}P},
\end{equation}
and,
\begin{equation}
\frac{dP}{dr}=-g\rho,
\end{equation}
where $g=GM_1 / r^2$. The structural polytropic index of the stellar profile is $\Gamma_{\rm s}= \left(\frac{d \ln P}{d \ln \rho} \right)_{\rm env}$. \\

The  gas envelope might have a different response to compression (as characterized by $\gamma_1$ and $\gamma_3$) than the one implied by the polytropic  index of the stellar profile. This is because rearrangements induced by the embedded object will happen on a timescale much shorter than the  thermal timescale of the evolving primary.  For example, a fully convective envelope might have $\Gamma_{\rm s} \approx \gamma_1$ whereas  a radiative envelope might  have $\Gamma_{\rm s} < \gamma_1$. 
In the case of an ideal gas, as considered in our FLASH calculations, we have the simplification $\gamma=\gamma_1=\gamma_3$. Regions where gas pressure dominates can be described by   a $\gamma=5/3$ while regions  where radiation pressure dominates are well characterized by a $\gamma=4/3$.

Locally, within this polytropic stellar envelope, the flow is described by dimensionless parameters such as the Mach number, 
\begin{equation}
\mathcal{M}=v_\infty / c_{s,\infty},
\end{equation}
and  the density gradient, 
\begin{equation}
\epsilon_{\rho}=R_{\rm a}/H_{\rho}.  
\end{equation}
Here  $H_{\rho}=-\rho dr/d\rho$, and $\epsilon_{\rho}$ represents the number of scale heights across  the accretion radius  within the primary
star's envelope, with $\epsilon_{\rho} \rightarrow 0$ describing an homogeneous density
structure and   $\epsilon_{\rho} \rightarrow \infty$ describing a very steep density
gradient. 
\citet{2017ApJ...838...56M} show that the expression 
\begin{equation}
\mathcal{M}^2=\epsilon_{\rho}\frac{(1+q)^2}{2q}f_{\rm k}^4\left(\frac{\Gamma_{\rm s}}{\gamma_1}\right),
\end{equation}
relates these flow parameters and describes pairings of $\mathcal{M}$ and $\epsilon_\rho$ for a given binary mass ratio and envelope structure. 

\subsubsection{Wind Tunnel Domain, Conditions, and Diagnostics}

The units of the simulations are such that $R_{\rm a}=v_{\infty}=\rho_{\infty}=1$, where $\rho_{\infty}$ is the density of the envelope at a distance $r = a$ from the primary's center. In this case, the characteristic time is  $t=R_{\rm a}/v_{\infty}=1$ and the mass of the embedded object is $M_2=(2G)^{-1}$. We employ $8\times5\times5$ initial blocks of $8^3$ cells in each direction, in a box of size $(-5,3)R_{\rm a}\times(-2.5,2.5)R_{\rm a}\times(-2.5,2.5)R_{\rm a}$. The maximum refinement level is set to 8, and the minimum is set to 2. Therefore, the maximum cell size is $R_{\rm a}/16$ and the minimum is $R_{\rm a}/1024$. The secondary is fixed at the origin and is surrounded by a sink boundary of radius $R_{\rm s}$.

As in the simulations of \citep{2017ApJ...838...56M}, the $-x$ boundary feeds a wind across the box in the $+x$ direction. The corresponding  gradient of 
pressure and density is constructed  in the $y$ direction and is uniform in the
$z$ direction. The conditions of the flow  are parametrized by a density gradient, $\epsilon_{\rho}$, an 
upstream Mach number, $\mathcal{M}$, and the
pressure and density at $y = 0$ given $q$, $f_{\rm k}$, $\Gamma_{\rm s}$ and $\gamma$.
Once the values at $y = 0$ are determined, the vertical structure of the flow ($\pm y$) is constructed using the equations of hydrostatic equilibrium. The structure of the flow is thus in hydrostatic
equilibrium with $M_1$'s gravitational force, which acts in the $-y$ direction.

To study the flow structure we employ $50^3$ passive particles as gas Lagrangian tracers, randomly distributed within  a rectangle of dimensions $(-3,-1)\times(-1,3)\times(-1,1)$ and evolved using a Runge Kutta scheme using the Particle module in FLASH \citep{2000ApJS..131..273F}. This technique allows us to study the capture and residence of fluid into rotational structures near the embedded object. This approach is valuable because the flow is highly time-variable, and, in these circumstances, single time snapshot streamlines can be misleading. 

\subsubsection{Simulation Parameters}

The key parameter that we vary across our simulations is the gas adiabatic index, $\gamma$. In so doing, we represent portions of the CE material with different compressibility, and, as we will show, different susceptibility to the formation of dense, rotationally-supported disk structures. 

The simulations adopt a sink boundary of $R_{\rm s} = 0.02 R_{\rm a}$, a central density gradient of $\epsilon_\rho=2$, a velocity fraction $f_{\rm k}=1$, and a mass ratio of $q=0.1$. These parameters may be compared to stellar envelope structures shown in \citet{2015ApJ...803...41M} and \citet{2017ApJ...838...56M}.
We will discuss the properties of typical stellar envelopes in Section~\ref{dis}, but for context, these conditions could represent those found in a $M_{\rm 1}=1 M_{\odot}$ giant branch star with  $R_1=140R_{\odot}$ engulfing a  $M_2=0.1 M_{\odot}$ star at a separation of $a=0.9 R_1$, or alternatively a $M_1=80 M_{\odot}$ red giant with  $R_{1}=740R_{\odot}$ engulfing a $M_2=8 M_{\odot}$ star at a separation of $a=0.85 R_1$.

We vary the adiabatic index across 4 simulations using (a) $\gamma=\Gamma_{\rm s}=5/3$, (b) $\gamma=\Gamma_{\rm s}=4/3$, (c) $\gamma=1.2$ with $\Gamma_{\rm s}=4/3$, and (d) $\gamma=1.1$ with $\Gamma_{\rm s}=4/3$.  We adopt $\Gamma_{\rm s} \ge 4/3$  in order to have a polytropic index that is stable to perturbations in pressure \citep{1958MNRAS.118..523B}.

The circularization radius depends on the density profile, and can be integrated numerically.  
We make use of $R_{\rm circ}=l_{z,\infty}^2/GM_2$, where 
\begin{equation}
l_{z,\infty}=\dot{L_z}(<R_{\rm a})/\dot{M}(<R_{\rm a}).
\end{equation} 
Numerical integration in the vertical direction can then provide 
\begin{equation}
\dot{M}(<R_{\rm a})=v_{\infty}\int_{<R_{\rm a}}\rho(y)\ dA,
\end{equation}
and 
\begin{equation}
\dot{L_z}(<R_{\rm a})=v_{\infty}^2\int_{<R_{\rm a}}\rho(y)y \ dA.
\end{equation}
The circularization radius therefore  depends solely  on the initial density profile, which in turn is set by $\Gamma_{\rm s}$ and $\epsilon_\rho$. 
For the conditions of our numerical simulations, with $\epsilon_\rho=2$, we find $R_{\rm circ}=0.35R_{\rm a}$ for $\Gamma_{\rm s}=4/3$  and $R_{\rm circ}=0.33R_{\rm a}$  for $\Gamma_{\rm s}=5/3$.

\section{Numerical Results}\label{res}

In  homogeneous HLA, where there is no density gradient, the gravity of the object focuses gas  into a stagnation region that trails behind it. Gas then flows into the object primarily in the opposite direction of the incoming material. The introduction of an upstream density gradient breaks the symmetry of the problem, altering the flow structure by introducing net angular momentum. Without the cancelation of momentum in the trailing stagnation region, the rate of mass accretion is drastically reduced when a gradient is introduced \citep{2015ApJ...803...41M,2015ApJ...798L..19M, 2017ApJ...838...56M}. 

\begin{figure}[tbp]
\begin{center}
\includegraphics[width=0.5\textwidth]{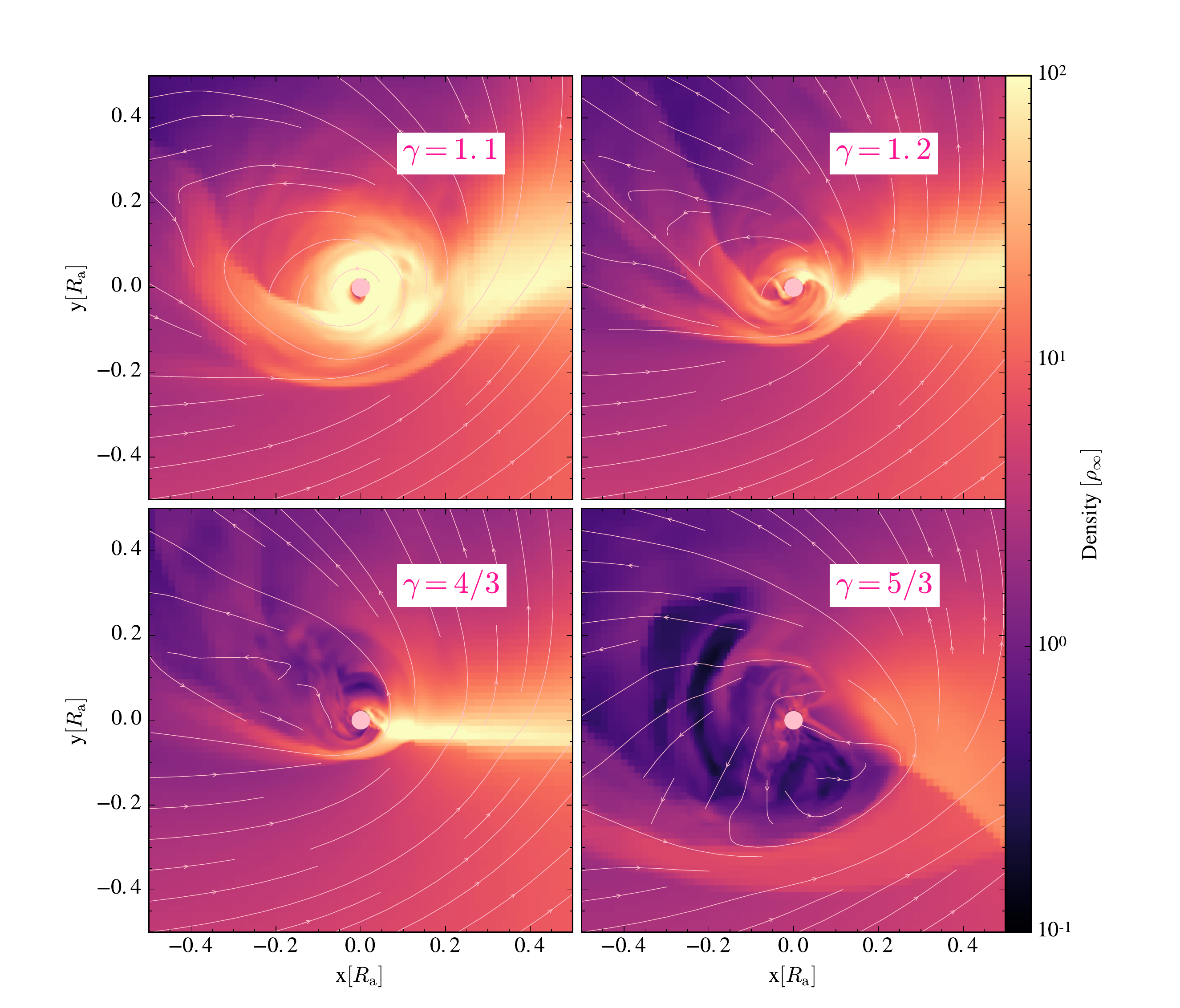}
\caption{Comparison of the flow morphologies in the orbital plane ($z=0$) with varying adiabatic indexes. All simulations are plotted at $t=25R_{\rm a}/v_\infty$. The simulation parameters are  $\epsilon_{\rho}=2$,$f_{\rm k}=1$, and $R_{\rm s}=0.02 R_{\rm a}$. The density has units of $\rho_\infty$. As can be seen, the density gradient tilts the shock, allowing for denser material to be deflected towards the outer edge, and the lower density material is more favorably accreted.  The streamlines show that in the lower $\gamma$ cases, a rotationally-supported structure can be formed. }
\label{fig:density_z}
\end{center}
\end{figure}

\begin{figure}[tbp]
\begin{center}
\includegraphics[width=0.5\textwidth]{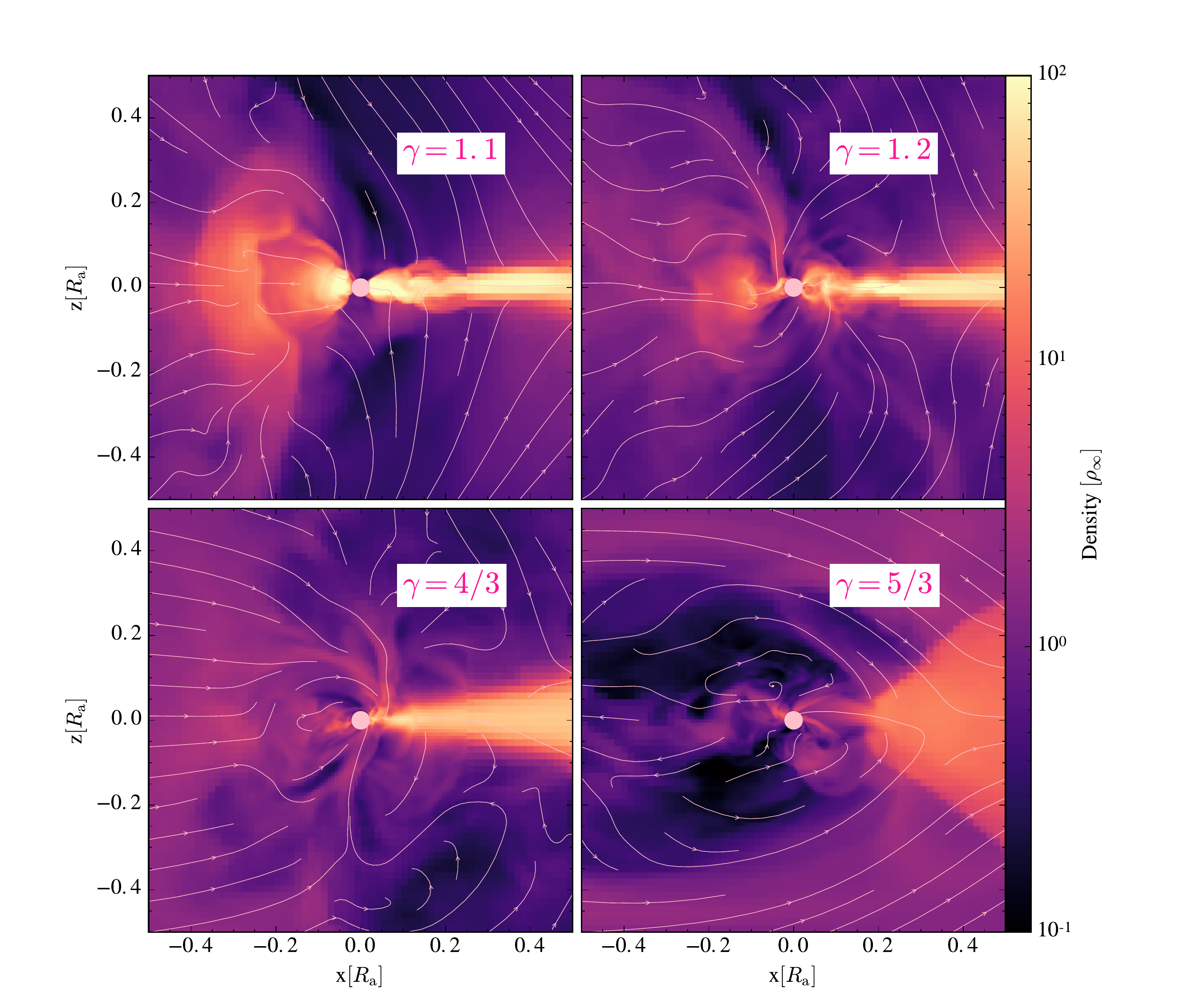}
\caption{Density and flow structure  for the same frames plotted in Figure~\ref{fig:density_z} but in the plane perpendicular to the orbit ($y=0$). The wake is more narrow and less dense in the less compressible media.  In all cases the flow is deflected toward the wake. In the $\gamma=5/3$ case, the streamlines are deflected away from  the accretor, creating cavities.}
\label{fig:density_y}
\end{center}
\end{figure}

Figure~\ref{fig:density_z} shows the structure of the flow in  the orbital plane for different adiabatic indexes. All simulation slices are plotted  at $t=25R_{\rm a}/v_\infty$. Due to the vertical density gradient, the incoming flow is preferentially deflected towards the lower density material located at the outer edges of the envelope. The flow lines clearly show that most of the dense  material, rather than being focused into the embedded object, is slingshotted into a counter-clockwise vortex. One or more angular momentum redistribution shocks form, which allow lower-density material to be  accreted more favorably by the embedded object.

A key property of the flow in our simulations is that there is a constant flux of new material flowing toward the accreting object. Interaction with this steady flow defines the structures seen in Figure~\ref{fig:density_z}. As the compressibility of the gas increases, there is an increase in the density near the accretor in order to maintain  ram pressure balance with the incoming material (with $P\propto \rho^\gamma$ along an adiabat, low $\gamma$ implies a need for high $\rho$ to match a pressure $P_{\rm ram} \approx \rho_\infty v_\infty^2$). The high densities near the accreting object imply that large quantities of material have pierced into the circularization region $r<R_{\rm circ}$.  Visually in Figure~\ref{fig:density_z}, we can see that the mass of material in the circularization region increases dramatically as we go from $\gamma=5/3$ to $\gamma=1.1$. Additionally, the centrifugal support of  gas near the accretor is  most prominent (both in streamlines and in density slice) when the flow is highly compressible.

We find that the ability for the incoming flow to settle into a dense, rotationally-supported disk depends sensitively  on the vertical structure of the flow, which is illustrated in Figure~\ref{fig:density_y}. Because of the varying thermal properties of the gas, the convergence region becomes narrower and more concentrated along the plane as the flow increases its compressibility. This enhanced vertical compression implies decreasing pressure relative to rotational support.\footnote{  We note that the convergence of  flow lines into a dense structure near the accretor leads the mass accretion rate to  increase with decreasing $\gamma$ by an order of magnitude between $\gamma=5/3$ and $\gamma=1.1$.  }

\begin{figure}[tbp]
\begin{center}
\includegraphics[width=0.5\textwidth]{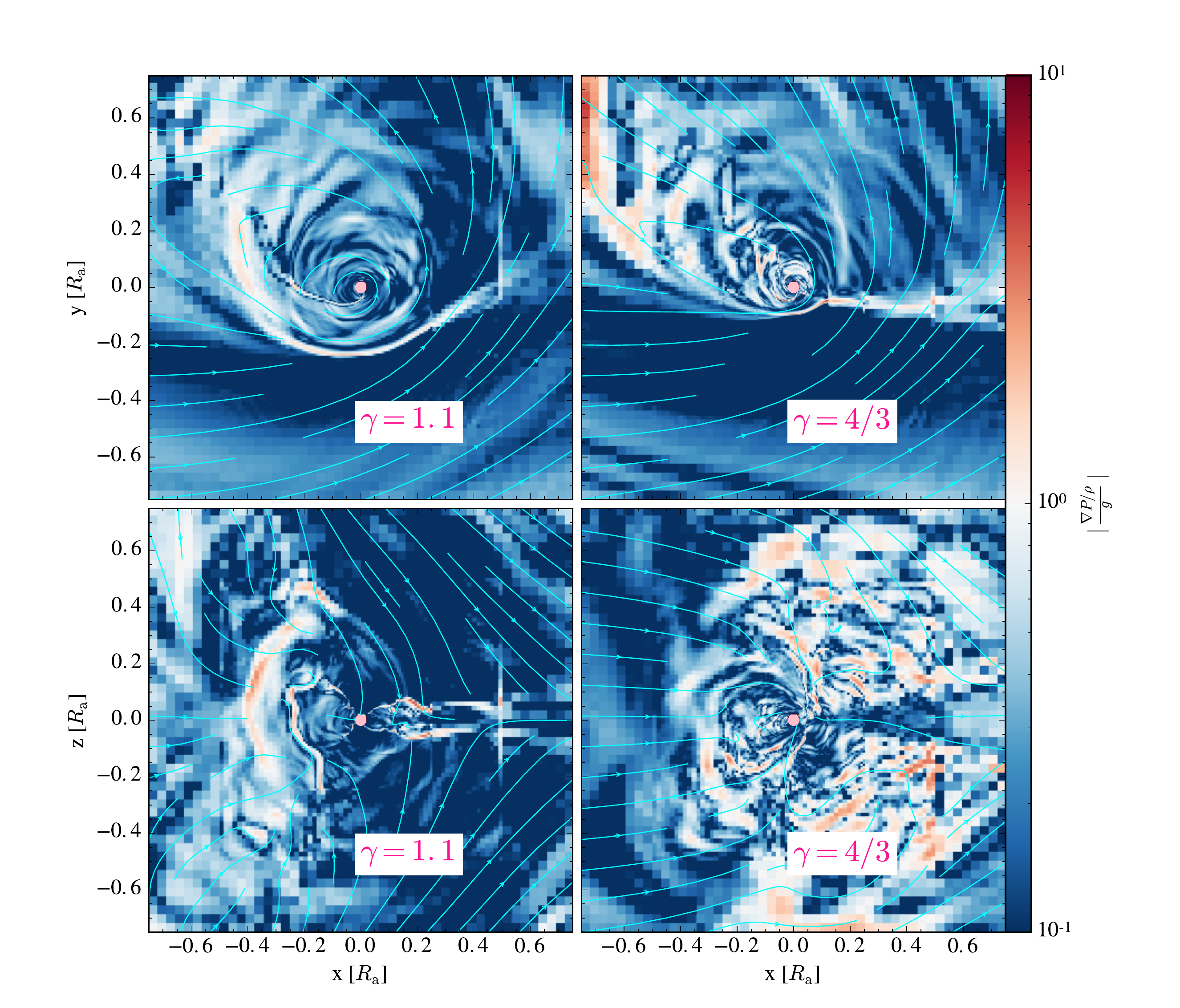}
\caption{Shown are the absolute value of the radial components of the pressure gradient over the gravitational force per unit density due to the embedded object ($g=GM_2/r^{2}$). The snapshots are the same as in Figures~\ref{fig:density_z} and \ref{fig:density_y}.  The  ratio of the forces in the orbital ({\it top} panel) and perpendicular  ({\it bottom} panel) planes are shown for  $\gamma=1.1$ and $\gamma=4/3$.   For the case of $\gamma=1.1$,  gravitational forces usually  dominate over the pressure gradient near the embedded object,  allowing the flow to reach the circularization region. In the $\gamma=4/3$ case, the pressure gradient dominates at large distances, which leads to stronger deflections of the flow. As a result, the flow is unable to enter the circularization region. }
\label{fig:pres_grad}
\end{center}
\end{figure}

To aid in understanding whether fluid is  able to approach the region of effective circularization,  in Figure~\ref{fig:pres_grad} we have plotted  the ratio of the absolute value of accelerations on the gas for  $\gamma=1.1$ and $\gamma=4/3$. 
In the highly compressible  case, we see that the gravitational force from the embedded object dominates over the pressure gradient  in most directions. This allows a sizable number of  flow lines to pierce into the circularization region without being substantially deflected by the collisional properties of the gas. For $\gamma=4/3$, on the other hand, the pressure gradient tends to dominate over the gravitational force and the flow is largely deflected away from the accretor.  Perpendicular to the orbital plane, the motion of the  adiabatic flow lines is influenced by the pressure gradient, thus leading to sizable defections of the flow away from the circularization region (with convergence happening primarily in the wake).  These deflections, as argued above, are less prominent in the $\gamma=1.1$ case, which allows  the gas to settle into a rotationally-supported structure. 

\begin{figure}[tbp]
\begin{center}
\includegraphics[width=0.47\textwidth]{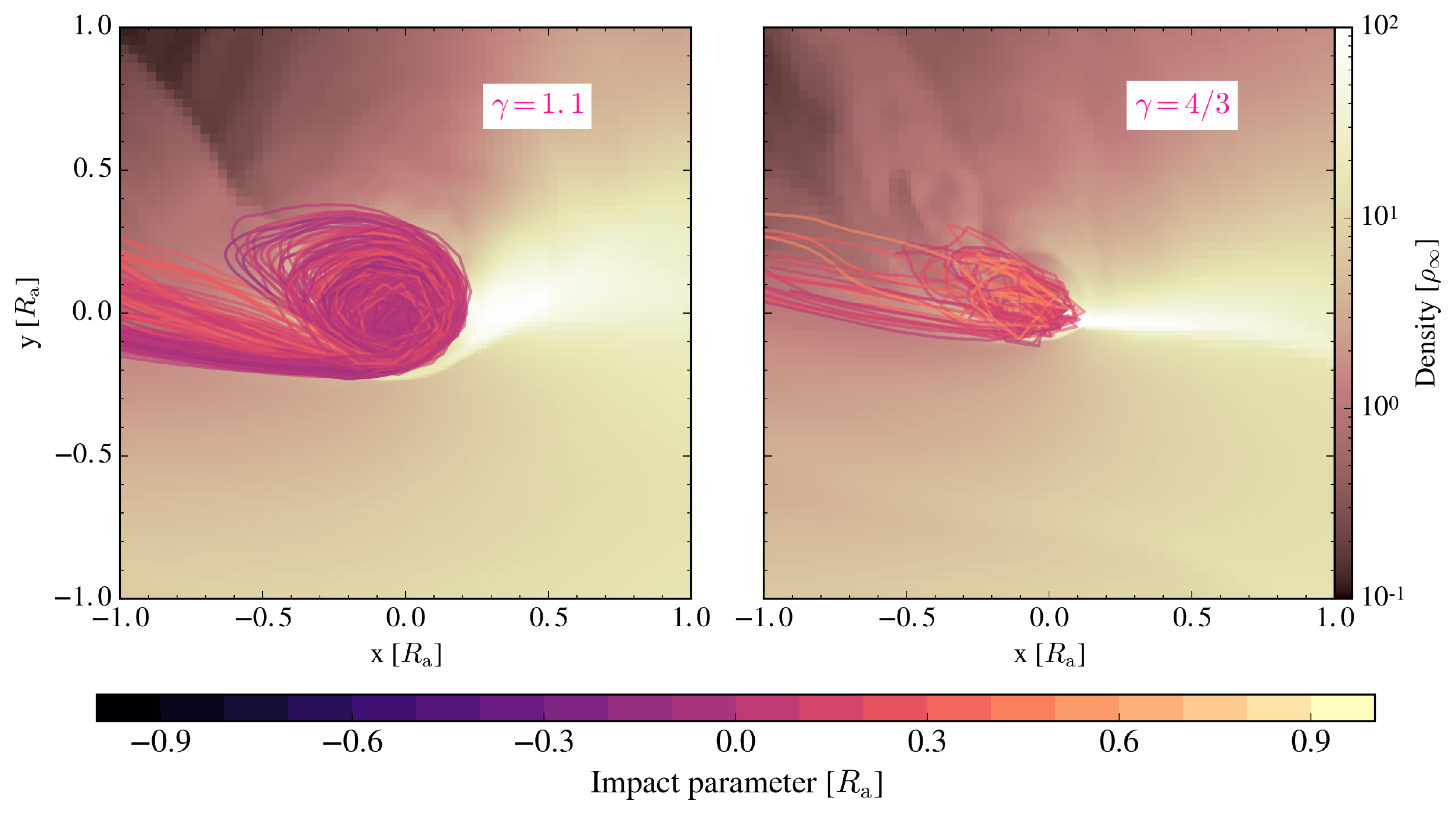}
\caption{Trajectories of 10\% of the injected  tracer particles (randomly selected)  belonging to the disk  region for an adiabatic index of $\gamma=1.1$ ({\it left} panel) and $\gamma=4/3$ ({\it right} panel).  We define particles that are part of the disk as particles that spend more than 15\% of the total time inside the circularization radius. Also shown is the initial impact parameter of each tracer particle and the density structure of the flow in units of $\rho_\infty$. }
\label{fig:trajectory}
\end{center}
\end{figure}

We explore the properties of this circularizing material further using our Lagrangian tracer particles of the simulation flow. 
Figure~\ref{fig:trajectory} selects particles that reside in the circularization region for more than 15\% of the time for which the particles are injected ($5 R_{\rm a}/v_\infty$). The trajectories plotted in Figure~\ref{fig:trajectory} are a randomly selected 10\% of those particles meeting the selection criteria.  Color indicates the initial impact parameters of the particles as injected into the domain.  For $\gamma=4/3$, a very small fraction of particles settle into the circularization region since most of them are deflected by the pressure gradient at larger distances.  A much larger number of tracer particles reside in the disk region when $\gamma=1.1$.  

Interestingly, Figure~\ref{fig:trajectory} also shows that fluid entering the circularization region in the $\gamma=1.1$ case originates almost entirely from impact parameters at or above the $y$-coordinate of the embedded, accreting object. In the context of the CE this corresponds to material at or outside the separation of the inspiralling object. The angular momentum redistribution shocks, coupled with the steep density gradient, appear to be the root of this behavior. In these shock structures angular momentum (relative to the embedded object) is transferred between fluid at positive and negative $y$ impact parameters.  The transfer is  preferentially from the higher angular momentum material to lower angular momentum material. Post-shock, material that has specific angular momentum capable to rotate at $r\approx R_{\rm circ}$ already interacted with the denser material and gained significant angular momentum.
The fact that material from $+y$ impact parameters has $+z$ angular momentum indicates that the direction of the angular momentum vector of these tracer particles was reversed as they passed through the redistribution shocks. 

\begin{figure}[tbp]
\begin{center}
\includegraphics[width=0.4\textwidth]{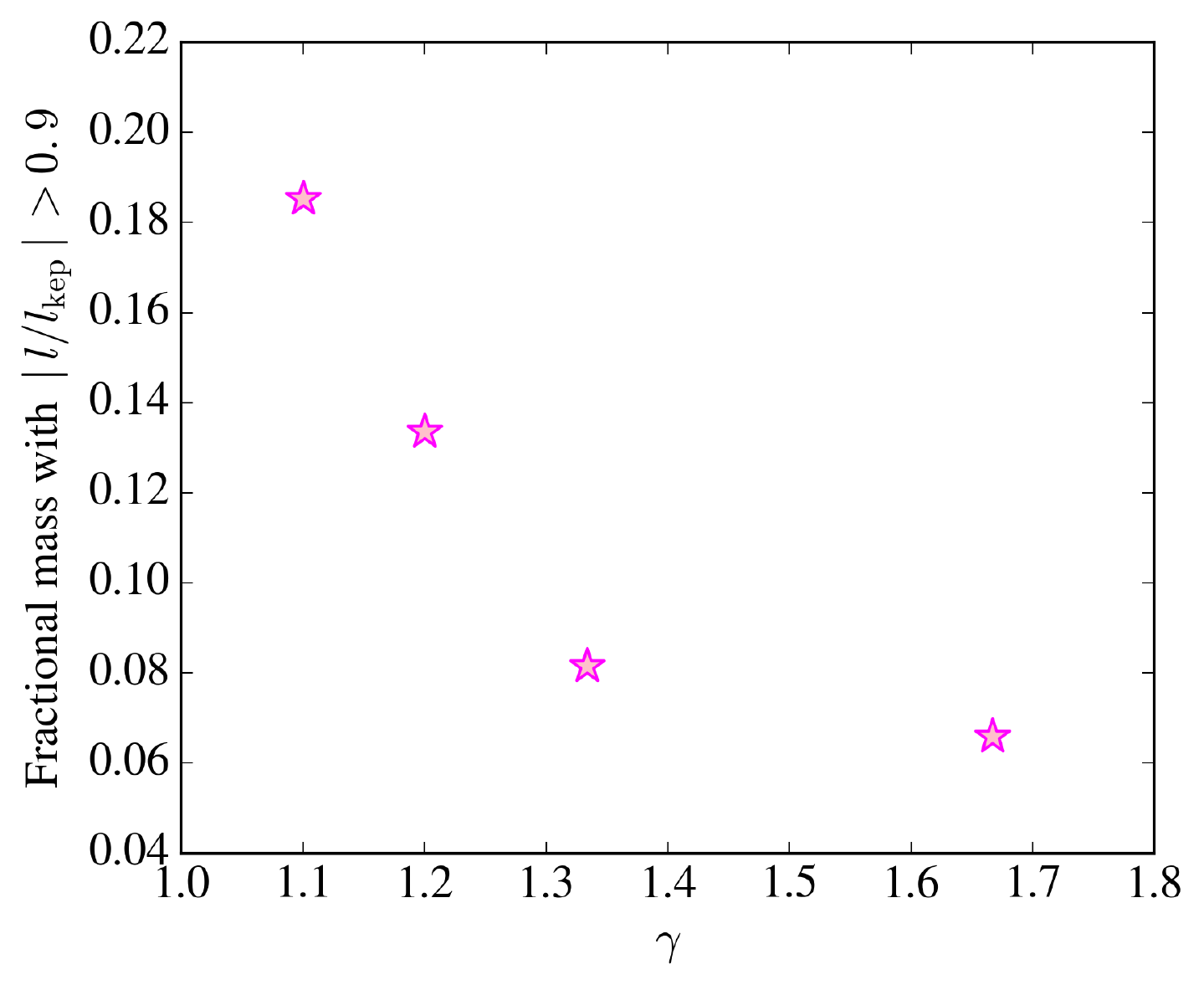}
\caption{Fractional mass inside the circularization radius and a height $z=0.1 R_{\rm a}$ with specific angular momentum $\mid l/l_{\rm kep}\mid>0.9$, where  $l_{\rm kep}=\sqrt{r/2}$ in our code units as a function of adiabatic index. The mass is normalized by the total mass inside the circularization radius. The total amount of mass with nearly Keplerian angular momentum decreases with decreasing compressibility, as the flow is unable to drill beyond $R_{\rm circ}$.}
\label{fig:mass_r_circ}
\end{center}
\end{figure}

The above analysis suggests that there is a critical adiabatic index below which a dense, rotationally-supported structure can be formed with these wind-tunnel flows.  Figure~\ref{fig:mass_r_circ} shows the total mass within the circularization radius that is rotationally-supported, defined here as having $\mid l/l_{\rm kep}\mid>0.9$. 
This highlights a conclusion which is visually obvious in Figures \ref{fig:density_z} and \ref{fig:density_y}: a highly compressible flow allows for a large amount of rotationally-supported material, a structure that we would typically consider a dense disk.  Figure~\ref{fig:trajectory} also shows that a relatively sharp transition  occurs below  $\gamma\approx 4/3$. In what follows,  we consider $\gamma\approx 1.2$ to be the representative critical value for  disk assembly, because our simulations with $\gamma\lesssim 1.2$ show disks, while those with $\gamma\gtrsim4/3$ do not.

\section{Discussion}\label{dis}

\subsection{Interpretation and Comparison to Previous Studies}
In this work, we have found centrifugally supported structures only for highly compressible flows $\gamma \lesssim 1.2$. This differs from \citet{2000ApJ...532..540A}, who reported disk formation in  radiation-dominated ($\gamma=4/3$) flows. The main reason for this discrepancy is undoubtedly the fact that they carried out simulations   in two  dimensions.   In three dimensions, the flow can be deflected  in the $z$-direction and is not restricted to the orbital plane. This additional  degree of freedom hinders disk formation \citep{1997A&A...317..793R}.  \citet{2015ApJ...803...41M} argued that pressure support under compression in two dimensions with an adiabatic  equation of state ($\gamma=5/3$) is very similar to that in a three-dimensional simulation with a nearly isothermal equation of state ($\gamma=1$). This is because $P \propto \rho^\gamma \propto V^{-\gamma}$, where $V$ is the volume term. In two dimensions we then have  $P_{\rm 2d} \propto r^{-2\gamma}$, while in three dimensions we can instead write $P_{\rm 3d} \propto r^{-3\gamma}$. As a result, $P_{\rm 2d} \propto r^{-10/3}$ ($P_{\rm 2d} \propto r^{-8/3}$) for  $\gamma=5/3$ ($\gamma=4/3$) and $P_{\rm 3d} \propto r^{-5}$ ($P_{\rm 3d} \propto r^{-3}$) for  $\gamma=5/3$ ($\gamma=1$).  

\begin{figure}
\begin{center}
\includegraphics[width=0.5\textwidth]{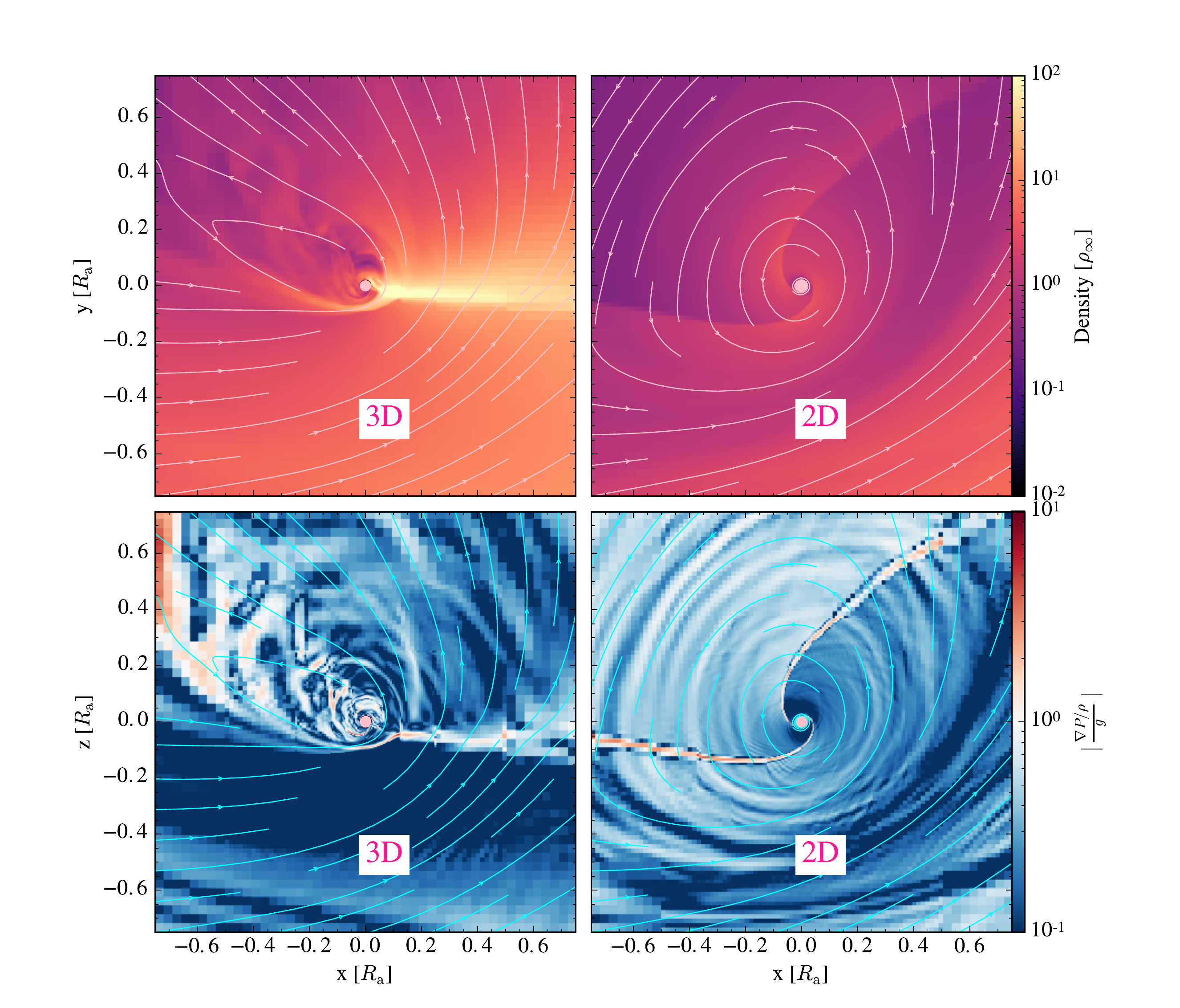}
\caption{Differences between a two-dimensional simulation and a three-dimensional simulation. Both simulations share the same initial conditions, $\gamma=\Gamma_{\rm s}=4/3$, $q=0.1$,$f_{\rm k}=1$, $\epsilon_{\rho}(y=0)=2$, $R_{\rm s}=0.02R_{\rm a}$. The flow is depicted in both cases  at a time $t=25 R_{\rm a}/v_\infty$. The initial number of blocks is $8\times5\times5$ (three dimensions) and $8\times5$ (two dimensions). The minimum refinement level is 2, and the maximum refinement level is 8 for both cases. Shown are  the density (\textit{top} panels) and the absolute value of the radial component of the pressure gradient over the gravitational force per unit density (\textit{bottom} panels), in the region near the accretor. The gravitational force dominates near the accretor in two dimensions, whereas the pressure support is significant in three dimensions. This results in a  rotationally-supported structure for two dimensions that is not present  in three dimensions.}
\label{fig:2dvs3d}
\end{center}
\end{figure}

Our analysis in Section~\ref{res} indicates that the radial component of the pressure gradient is more important than the pressure itself, because this is the quantity that enters into the gas momentum equation, as $\nabla P/\rho$. 
If we consider the idealized case of spherical compression, the pressure gradient term, in three dimensions with a nearly isothermal equation of state or in two dimensions with an adiabatic one,  $\frac{1}{\rho}\frac{dP}{dr}\propto r^{-1}$. By contrast,  the pressure gradient in three dimensions is $\frac{1}{\rho}\frac{dP}{dr}\propto r^{-3}$ for an adiabatic flow. Thus, near the embedded object, the resistance to compression due to the pressure gradient  is much stronger for the adiabatic case than for the isothermal one. Figure~\ref{fig:2dvs3d} compares simulations in three and two dimensions with $\gamma=4/3$. The flow in  two dimensions is rotationally-supported, as also seen by \citet{2000ApJ...532..540A}, while the  increase in pressure support  in three dimensions does not allow the flow to circularize.

A sufficiently strong pressure gradient can act effectively against the gravitational force of the embedded object, which goes as $\propto r^{-2}$. Returning to our three-dimensional flow structures, this leads to larger deflections of  the flow in the $\gamma=5/3$ case, as observed in Figures~\ref{fig:density_z} and \ref{fig:density_y}, which prevent  the  formation of a dense disk. If the resistance of a pressure gradient against gravity is the controlling parameter, we find that these both scale as $r^{-2}$ for $\gamma=4/3$, implying that in initial ratio of pressure support to gravitational acceleration is preserved at all radii under spherical compression. To settle into a disk, we can imagine that fluid needs to have a pressure-gradient scaling shallower than $r^{-2}$, so that gravity can become dominant at some radii and a rotationally-supported flow can develop. This logic predicts a bifurcation in  flow structure  above and below $\gamma \approx 4/3$. Our results of Section~\ref{res} support that prediction: only in calculations with $\gamma < 4/3$ ($\gamma \lesssim 1.2$) did we find dense disks on the scale of $R_{\rm circ}$. 

\subsection{Where in CE inspiral can disks form?}

 \begin{figure}
\begin{center}
\includegraphics[width=0.42\textwidth]{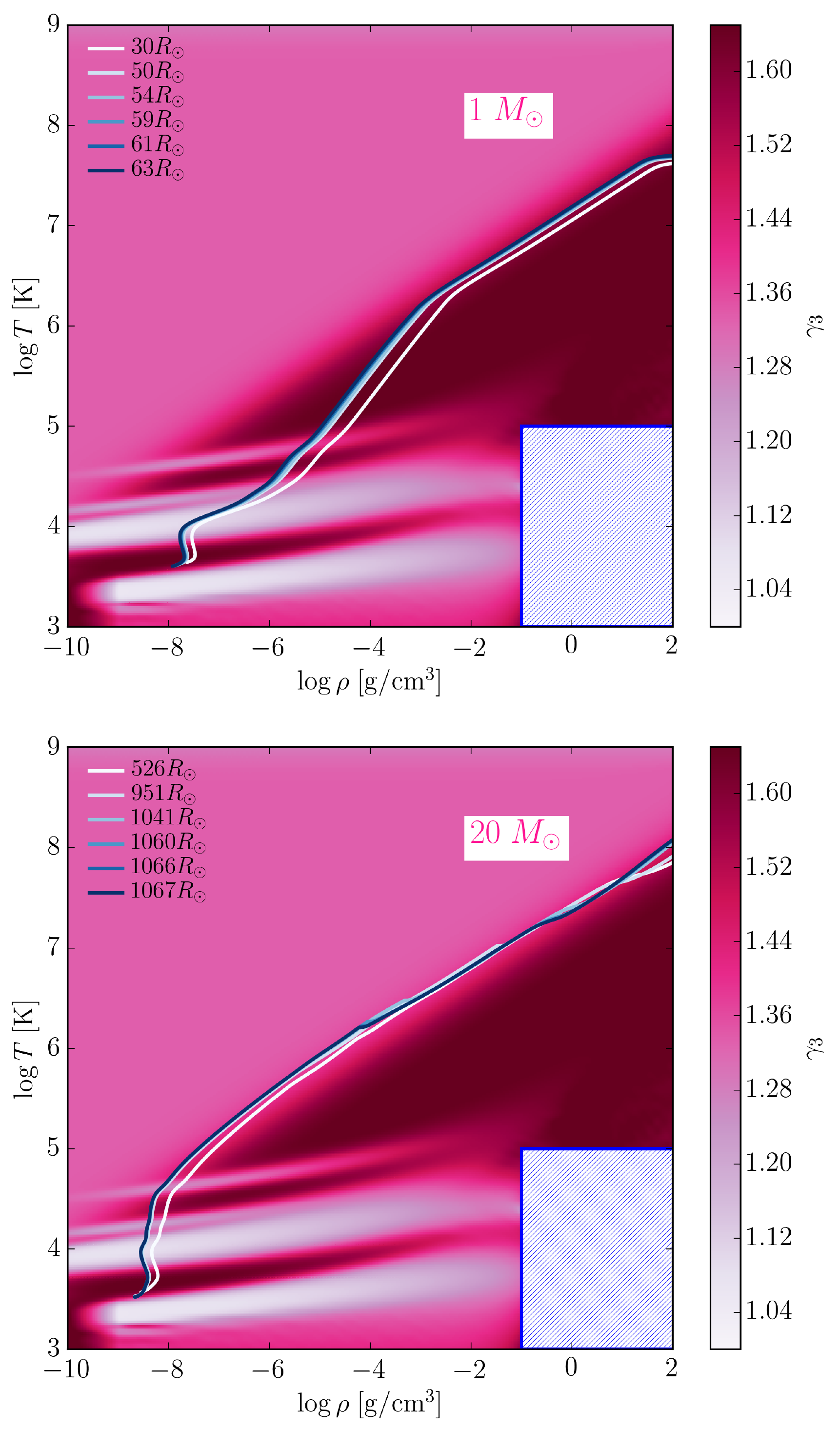}
\caption{Mapping of $\gamma_3$ in stars with solar abundances. Over-plotted are the tracks of stars with $M=1 M_{\odot}$ ({\it top} panel), and $M_1=20 M_{\odot}$ ({\it bottom} panel) for different evolutionary  stages. The dashed area represents a region in the $\rho-T$ plane where crystallization occurs and  the equation of state is not well determined \citep{2011ApJS..192....3P}. The regions with low adiabatic index correspond to partial ionization zones \citep{1984MNRAS.210..633H}.  In most stars there are two main ionization zones. The hydrogen partial ionization zone where both the ionization of neutral hydrogen ${\rm H} \leftrightarrow {\rm H}^+ + {\rm e}^-$ and the first ionization of helium  ${\rm He} \leftrightarrow {\rm He}^+ + {\rm e}^-$ occurs in layers with a characteristic temperature of $1.5 \times 10^4$ K. The second involves the second ionization of helium ${\rm He}^+ \leftrightarrow {\rm He}^{++} + {\rm e}^-$, which occurs in deeper layers with  a characteristic temperature of $4 \times 10^4$ K. Upon compression, internal 
energy is partially deposited into increased ionization within these regions, lowering $\gamma_3$.}
\label{fig:mesa_eos}
\end{center}
\end{figure}

As discussed previously, during a CE event, a rotationally-supported structure  could form around the embedded object in the presence of highly compressible gas. Natural questions then arise: where in a stellar envelope can this occur? And what is the scale of the associated disk? 

The highest compressibility environment found in stars is within partial ionization zones.  In these zones where the gas is partially ionized, a fraction  of the energy released during a layer's compression can be used for further ionization, rather than raising the temperature of the gas \citep{1984MNRAS.210..633H}.  The partial ionization produces an opacity bump and a considerable decrease in the adiabatic exponents. As a result, a steeper temperature gradient is required  in order for  radiative diffusion to   transport energy through these regions.  

Such partial ionization zones are  located  in the outer layers of evolving  stars. To illustrate  this, we calculate stellar models with MESA \citep[version 7624;][]{2011ApJS..192....3P, 2013ApJS..208....4P, 2015ApJS..220...15P}  for stars of different mass and evolutionary stages \footnote{We evolved the stars with masses $M_1=15, 20, 30,40,50,60,70,80 M_\odot$ using the   \texttt{150M\_z1m4\_pre\_ms\_to\_collapse} test suite setup, but changing the initial mass and metallicity accordingly. The setup does not alter the \texttt{inlist\_massive\_defaults}, which includes a mixing length of 1.5 and a `Dutch' wind scheme for both RGB and AGB winds. The stars with masses $M_1=1,2,5,10 M_\odot$ were evolved using the setup from the test suite  \texttt{7M\_prems\_to\_AGB}, but, again, changing the masses accordingly. The setup uses a mixing length of 1.73, and a `Reimers' and `Blocker' RGB and AGB wind schemes respectively. The corresponding  inlists are available upon request.}.

As the star evolves into the giant branch, the partial ionization regions occupy a progressively larger fraction of the mass of the star. This can be seen in Figure~\ref{fig:mesa_eos}, where we have mapped the compressibility that enters into the equation of state, $\gamma_3$ (equation \ref{eos}), in the $\rho-T$ plane  using the equation of state module in MESA \citep{2011ApJS..192....3P, 2013ApJS..208....4P, 2015ApJS..220...15P}. Over-plotted are the evolutionary tracks for  $1 M_{\odot}$ stars and $20 M_{\odot}$ stars at various evolutionary stages and solar abundance.  The dashed area represents the region where crystallization occurs and the equation of state is not well determined. 

The almost horizontal (constant $T$) white bands in  Figure~\ref{fig:mesa_eos} represent  regions in which partial ionization of various species takes place and, as a result, the gas is highly compressible. The regions of high compressibility  are more prominent in low mass stars, whose envelopes cross through larger portions of these regions. High mass giants approach their Eddington limit and show profiles in the $\rho-T$ plane that straddle the gas-radiation pressure transition. These profiles touch the partial ionization  regions in $\rho-T$ space only at their extreme limbs, occasionally in regions of density inversion. 

\begin{figure}
\begin{center}
\includegraphics[width=0.39\textwidth]{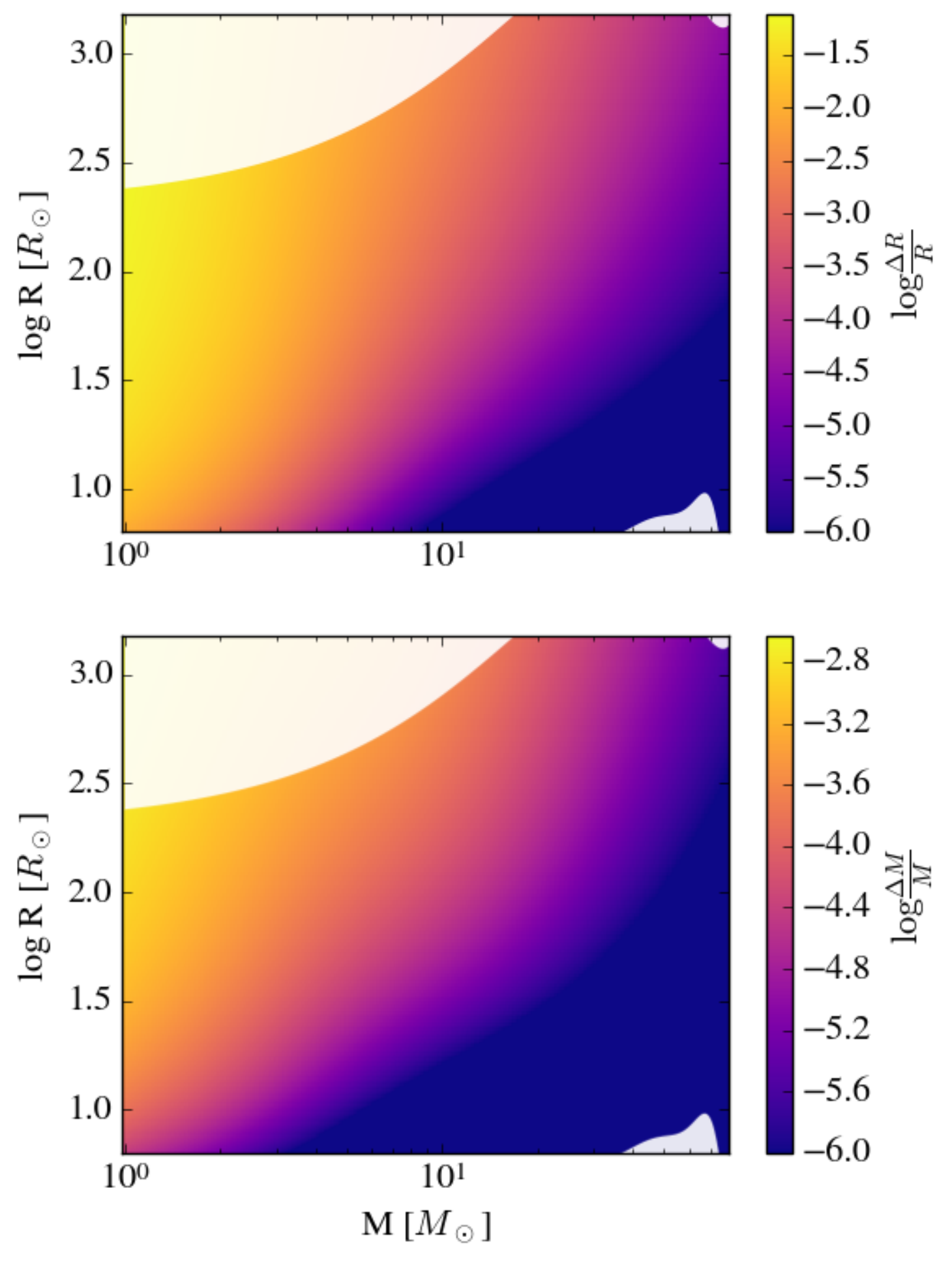}
\caption{Fractional radius (\textit{top} panel) and fractional mass (\textit{bottom} panel) as a function of initial mass and radius of solar metallicity stars having a high compressibility zone ($\gamma_3<1.2$). 
The fraction of the stars having a highly compressible region is significantly  more extended in low mass stars compared to high mass stars. }
\label{fig:fraction_adiabatic}
\end{center}
\end{figure}

Figure~\ref{fig:fraction_adiabatic} shows the fractional radius of  stars  that have $\gamma_3<1.2$. 
As can be clearly seen in Figure~\ref{fig:fraction_adiabatic}, low mass stars have significantly more extended  partial ionization zones in their outer layers.  Higher mass stars, above $\approx 3 M_\odot$ exhibit radially narrow partial ionization zones with $\lesssim 1\%$ of their radius occupied by these regions. 

\begin{figure}[tbp]
\begin{center}
\includegraphics[width=0.5\textwidth]{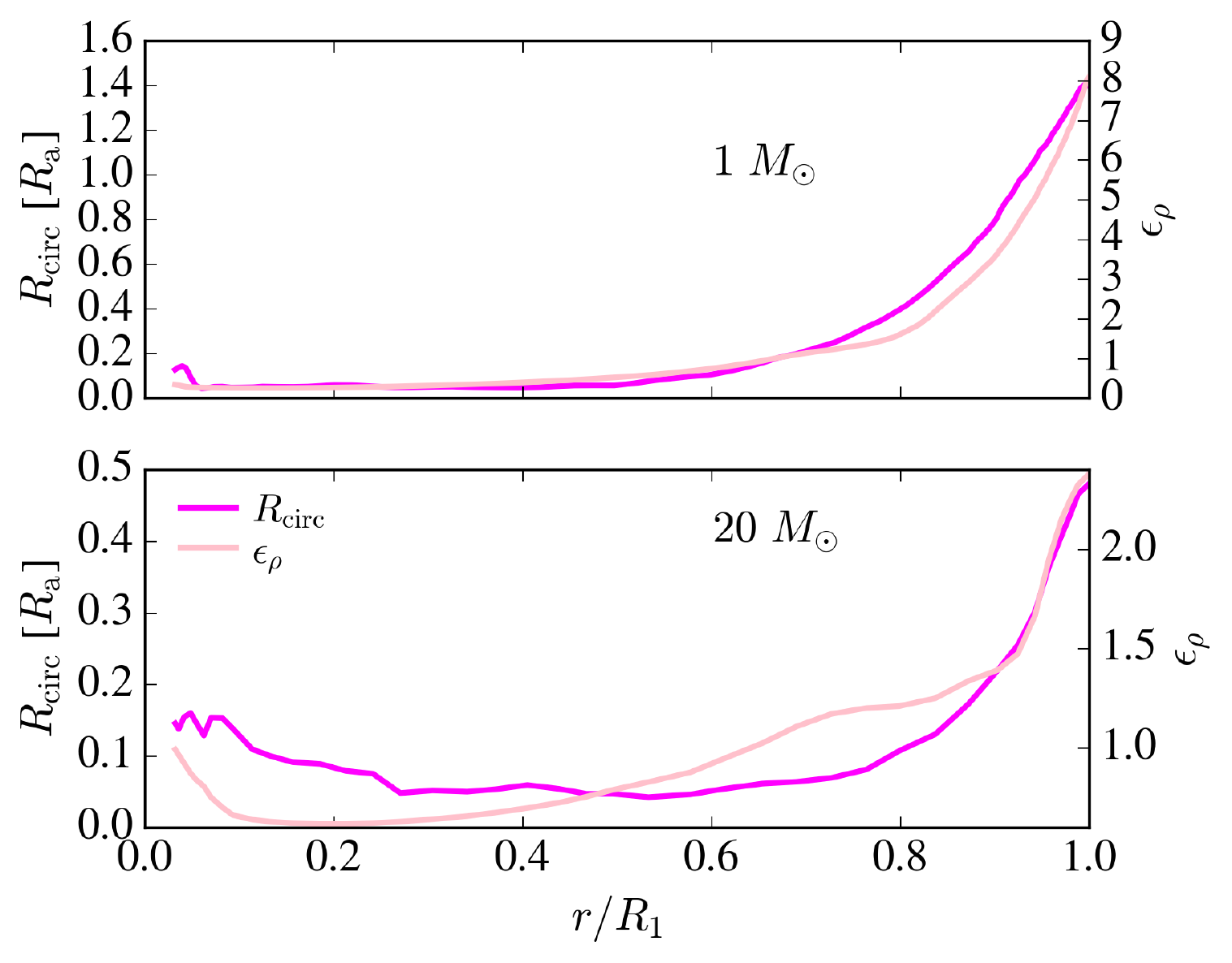}
\caption{Circularization radius and density gradient as a function of stellar radius for a   $1\ M_{\odot}$ primary star  with $R_1=30\ R_{\odot}$ at $6.3\times 10^9$ years and a helium core of $0.35 M_\odot$ (\textit{top} panel), and a $20\ M_{\odot}$ primary star with  $R_1=1000 \ R_{\odot}$ at $8.1\times10^6$ years and a helium core of $4.9 M_\odot$ (\textit{bottom} panel). Both stars have solar metallicity, and the binary mass ratio is assumed to be $q=0.1$. This figure was made using MESA \citep[version 7624;][]{2011ApJS..192....3P, 2013ApJS..208....4P, 2015ApJS..220...15P}. The circularization radius, as a fraction of $R_{\rm a}$, is highly dependent on the local density gradient. It is similar to $R_{\rm a}$ in the outer portion of the stellar envelope, but then decreases to $R_{\rm circ}\sim 0.1 R_{\rm a}$ for much of $r/R$.} 
\label{fig:r_circ}
\end{center}
\end{figure}

Next we address the scale of a disk that might result from passage of a secondary object through one of these regions of high gas compressibility. Figure~\ref{fig:r_circ} illustrates how the gas circularization radius, $R_{\rm circ}$, changes as the embedded object, here characterized by  $R_{\rm a}$, spirals deeper   into the star. This figure adopts $q=0.1$. As the embedded companion spirals deeper into the primary, density gradients, as parameterized by $\epsilon_\rho$, become shallower, and the circularization radius decreases relative to $R_{\rm a}$. Rotationally supported structures will have scale similar to $R_{\rm a}$ only in the outer portions of stellar envelopes, in similar regions to where zones of partial ionization (and high compressibility) are found.

We expect that only in cases where the radial extent of highly compressible gas ($\gamma < 4/3$) is sufficiently large, $\Delta R_1 \gtrsim R_{\rm a}$, is it possible for a large scale disk structure at the $R_{\rm circ}$ scale to be formed around the embedded object.  We have argued that such a disk would transport mass and energy to small scales from which it might generate accretion-driven winds and collimated outflows \citep{2000ApJ...532..540A,2003MNRAS.342.1169V, 2004NewA....9..399S,2015ApJ...800..114S}.  This figure indicates that, given the unperturbed structures of stellar envelopes, disk assembly and the corresponding accretion feedback during CE might be restricted to objects embedded in the outer envelopes of low-mass giant  stars.  Furthermore, by comparison to Figure~\ref{fig:fraction_adiabatic}, for $R_{\rm a} \lesssim \Delta R$, to occur, the encounter must be one with a low mass secondary object and correspondingly low mass ratio, $q$, such that $R_{\rm a} \ll a$. Taken together, these considerations suggest that disk formation in CE is rare, and is probably only a brief phase during the inspiral in cases in which it does occur. 

Several caveats affect the firmness with which we can reach such a conclusion. CE structures are undoubtedly expanded by interaction with the secondary star, perhaps even prior to the phase when an object plunges through a given radial coordinate. This expansion, and associated adiabatic degradation of the temperature of the expanded envelope, could lead larger portions of the $\rho-T$ trajectories of massive stars to cross through partial ionization zones. A second concern relates to the extension of our wind-tunnel results to the realistic CE process. In particular, in a full equation of state, such as that shown in Figure~\ref{fig:mesa_eos}, $\gamma_3$ is a strong function of density and, especially, temperature. This might lead to different structures (and degrees of pressure support) as the gas compresses through various phase transitions, perhaps differentiating the dynamics of the system under a realistic equation of state from that with a constant $\gamma$.  For now, we can speculate that the important scale is the circularization radius scale, where the angular momentum budget is dominated, but performing more complex simulations is beyond the scope of the current work. 

\subsection{Implications for binary black holes}
\begin{figure}[tbp]
\begin{center}
\includegraphics[width=0.42\textwidth]{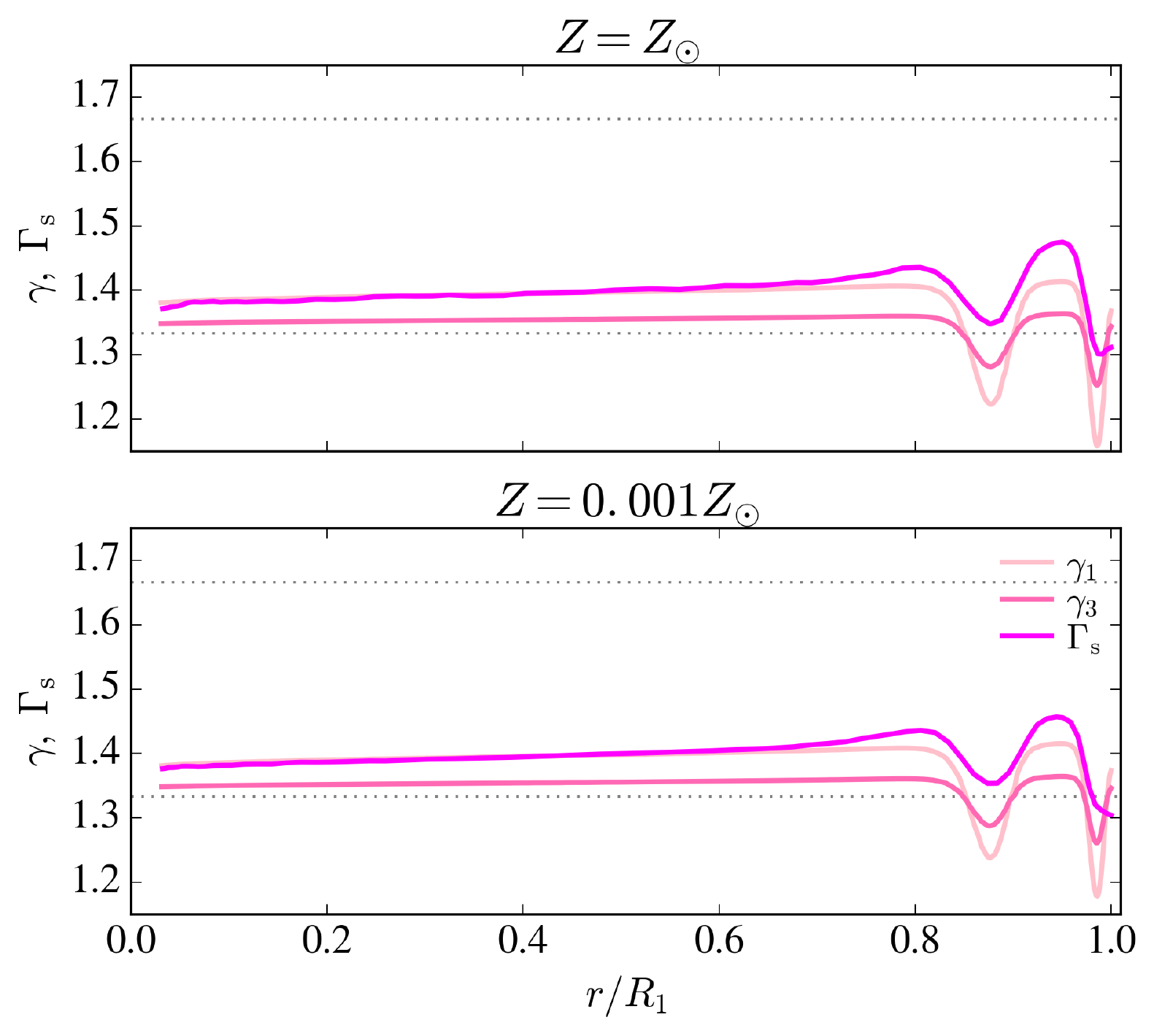}

\caption{Adiabatic indexes $\gamma_1$, $\gamma_3$, and structural index $\Gamma_{\rm s}$ in a $M_1=80 M_{\odot}$ star.  The {\it top}  and {\it bottom} panels shows the profiles of stars  with $Z=Z_{\odot}$ with $R_1=667 R_\odot$ and  $Z=0.001 Z_{\odot}$ with $R_1=667 R_\odot$, respectively. The stars are in an evolutionary stage where, for the solar metallicity star the helium core has a mass of $35 M_\odot$ and age of $3.2\times 10^6$ years, and the lower metallicity star has a helium core of mass of $40 M_\odot$ and age of $3.5\times 10^6$ years. Stars were evolved using MESA's inlist  \citep{2011ApJS..192....3P, 2013ApJS..208....4P, 2015ApJS..220...15P} \texttt{150M\_z1m4\_pre\_ms\_to\_collapse} test suite setup, but changing the mass, and metallicity accordingly.  As can be seen, the partial ionization regions (where $\gamma_1$ and $\gamma_3$ dip) are narrow and independent of metallicity. }
\label{fig:mesa_eos_low_met}
\end{center}
\end{figure}

The first detection of gravitational waves was catalyzed  by the existence of moderately massive, stellar-mass black holes in binary systems \citep{2016PhRvL.116f1102A}. One of the preferred channels for the formation of this type of binary black holes necessitates  a CE  stage \citep{2016A&A...596A..58K,2016Natur.534..512B,2012ApJ...759...52D,2016Natur.534..512B}. This channel involves a massive stellar binary ($40-100 M_{\odot}$), likely formed in a low-metallicity environment, in which the first-born black hole  is engulfed by an evolving massive companion \citep{2016Natur.534..512B}. For the merger to occur within the age of the universe, the black hole needs to  tighten its  orbit before the CE is ejected. 

We argue here that  accretion feedback is likely not to effectively operate  during  the CE phase when involving massive stars. As was shown in Figures~\ref{fig:mesa_eos} and \ref{fig:fraction_adiabatic}, extended zones of sufficiently high gas compressibility for disk formation exist in the envelopes of low mass giants, and are found in  zones of partial ionization. The extent of these zones is drastically reduced and they are only found  in the outermost envelopes of the high mass stars that are relevant to the formation of binary black holes.  In Figure~\ref{fig:mesa_eos_low_met} we additionally illustrate that this conclusion is not sensitive to varying  metallicity. 

Because no extended regions of sufficiently low $\gamma$ exist to allow disks to form in CE events involving high mass giants, there is a lack of a mechanism (such as a disk outflow) to couple the accretion energy lost to an embedded black hole with the large scale flow. 
This implies that significant tightening of the orbit  can take place without significant feedback energy injection from the embedded black hole. 
By avoiding strong feedback, CE events may serve as a mechanism to drive substantial orbital tightening, as originally envisioned \citep{1984ApJ...277..355W}, and are a natural channel to the formation of merging binary black holes \citep[e.g.][]{2016Natur.534..512B,2016A&A...596A..58K}.

\subsection{Summary}\label{conclusion}
In this paper, we study the conditions required  to form a disk around  the embedded companion during a CE phase. We studied the flows using the idealized CEWT setup of \citet{2017ApJ...838...56M}.  Some key conclusions of our study  are:
\begin{enumerate}

\item The introduction of a density gradient in HLA allows for angular momentum to be introduced to the flow, which in turn opens the possibility for the  formation of a disk around the embedded companion.

\item The formation of disk structures in the context of a CE phase is linked to the thermal properties of the envelope. In envelope gas with higher compressibility $(\gamma < 4/3)$, the gravitational force dominates over the pressure support near the accretor,  allowing for effective  circularization of the material into a disk. On the other hand, in  lower compressibility gas environments $(\gamma \gtrsim 4/3)$, the pressure support dominates as the gas compresses toward the accretor. We find that a disk does not form and the flow will be advected away from the embedded object, typically completing less than one full rotation.
 
\item Within stellar envelopes extended regions of sufficiently compressible gas to allow disk formation around embedded objects are found only within zones of partial ionization, where the additional (ionization) degrees of freedom reduce $\gamma$ significantly. 

\item These partial ionization zones always comprise a small fraction of a stellar envelope radius or mass. They are more extended in the outer layers of low mass stars than in the exteriors of high mass stars. We therefore expect that disk formation around embedded objects in CE, is, at most, a transitory phase. 

\item The lack of regions conducive to disk formation in high-mass stellar envelopes suggests that CE episodes involving these stars, such as those in the assembly history of merging binary black holes, are not subject to strong disk-outflow powered accretion feedback. Without overwhelming feedback from accretion, we suggest that CE events in massive systems should proceed with significant orbital tightening as they draw on orbital energy as an CE ejection mechanism rather than accretion energy. The lack of feedback implies that CE events remain a natural channel for the formation of LIGO-source binaries that must be assembled into tight orbits from which they merge under the influence of gravitational radiation. 
\end{enumerate}

\section*{Acknowledgements }
We thank S. de Mink, R. Foley, E. Quataert, E. Gentry, J. Law-Smith, R. Murray-Clay, J. Schwab, and M. Zaldarriaga for insightful discussions. The software used in this work was in part
developed by the DOE-supported ASCI/Alliance Center for
Astrophysical Thermonuclear Flashes at the University of
Chicago. For the analysis, we used {\it yt} analysis toolkit \citep{2011ApJS..192....9T}. 
 This research made use of {\tt astropy}, a community developed
core Python package for Astronomy \citep{2013A&A...558A..33A}. 
The calculations for this research were carried out in part on the UCSC supercomputer Hyades, which is supported by the National Science Foundation (award number AST-1229745) and UCSC. 
A.M.B. acknowledges UCMEXUS-CONACYT Doctoral Fellowship.
M.M. is grateful for support for this work provided by NASA through Einstein Postdoctoral Fellowship grant number PF6-170155 awarded by the Chandra X-ray Center, which is operated by the Smithsonian Astrophysical Observatory for NASA under contract NAS8-03060.
A.A. gratefully acknowledges support from the NSF REU program LAMAT at UCSC, a UCSC Undergraduate Research in the Sciences Award, and the California Space Grant Consortium (CaSGC) Undergraduate Research Opportunity Program.
P.M. is supported by an NSF Graduate Research Fellowship and a Eugene Cota-Robles Graduate Fellowship. E.R.-R. acknowledges financial support from the Packard Foundation and NASA ATP grant NNX14AH37G.
Additional support for this work is provided through program HST-AR-14574.002-A by NASA through a grant from the Space Telescope Science Institute, which is operated by the Association of Universities for Research in Astronomy, Inc., under NASA contract NAS 5-26555.
\bibliographystyle{aasjournal}
\bibliography{bibliography}

\end{document}